\documentstyle[prl,twocolumn,aps,epsf]{revtex}
\newcommand{\be}{\begin{equation}}
\newcommand{\ee}{\end{equation}}

\begin{document}

\twocolumn[\hsize\textwidth\columnwidth\hsize\csname @twocolumnfalse\endcsname
\draft
\tolerance 5000

\title{Transmission through quantum networks}

\author{Julien Vidal$^{1}$,  Gilles Montambaux$^{1}$ and  Benoit Dou{\c
c}ot$^{2,1}$}

\address{$^1$ Laboratoire de Physique des Solides, CNRS UMR 8502, Universit\'e
Paris Sud, B{\^a}t. 510, 91405 Orsay, France}

\address{$^2$ Laboratoire de Physique Th\'{e}orique et Hautes \'Energies,
CNRS UMR
7589, Universit\'{e}s Paris 6 et 7,\\ 4, place Jussieu, 75252 Paris Cedex 05
France}
\maketitle

\begin{abstract} {
We propose a simple formalism to calculate the conductance of any quantum
network
made  of one-dimensional quantum wires.
We apply this method to analyze, for two periodic systems, the modulation
of this conductance
with respect to the magnetic field.
We also study the influence of an elastic disorder on the periodicity of
the AB oscillations,
and we show that a recently proposed localization mechanism induced by the
magnetic field
resists to such a perturbation.
Finally, we discuss the relevance of this approach for the understanding of a
recent experiment on  GaAs/GaAlAs networks.
}
\end{abstract}

]

It is well  known that  quantum transport exhibits deviations from
classical transport, resulting in corrections to the classical addition
rules of
conductances or resistances. A spectacular example
 is the Aharonov-Bohm (AB) effect, where the  conductance of a ring is a
periodic
function of the  magnetic flux $\phi$ through its opening, with period
$\phi_0=h/e$. Since the first observation of  this effect in condensed
matter\cite{Webb}, many papers have been devoted to the study of coherence
effects
in transport, especially in the  ring geometry. A first approach uses the
Landauer
formalism in which the conductance   is proportional to the transmission
coefficient. In this framework, disorder effects have been considered in
single-channel \cite{Buttiker1} and multi-channel rings
\cite{Buttiker2}. On the other hand, the conductance of
 diffusive systems has been also extensively studied within the  Kubo approach,
where the  weak-localization correction is related to the modulation by the
magnetic field of  the return probability of a diffusive particule
\cite{Aronov87}. Although being a transport property, this correction is a
spectral
quantity, since it is related to the spectrum of the diffusion equation,  more
precisely to its spectral determinant \cite{Pascaud99}.

\begin{figure}
\centerline{ \epsfxsize=80mm
\epsffile{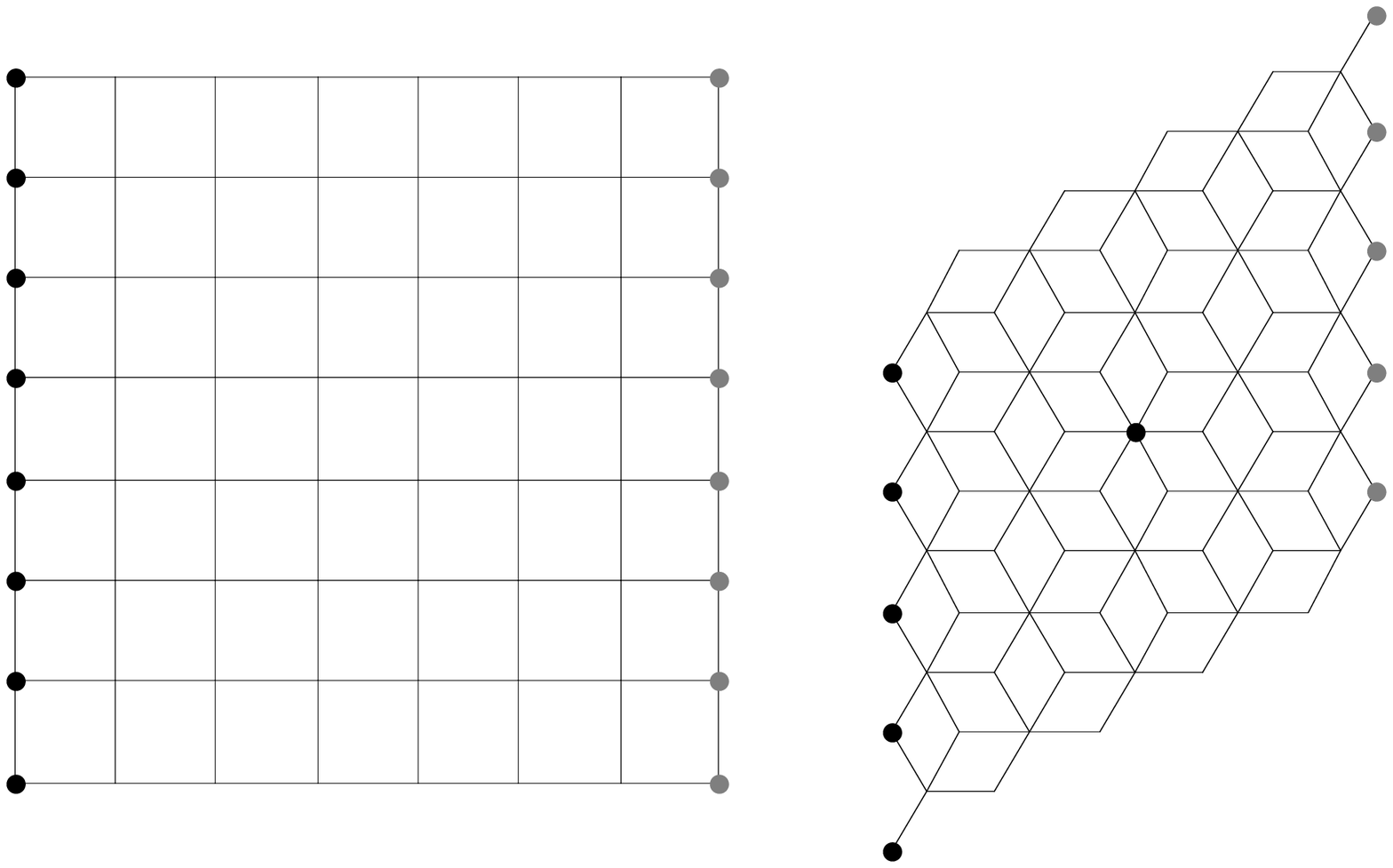}}
\vspace{3mm}
\caption{A piece of the ${\cal T}_3$ network (right) and of the square
lattice (left).
Black (respectively grey) dots represent the connections to the input
(respectively output)
channels. The central black dot is the input  channel chosen for the bulk
injection. }
\label{figure1}
\end{figure}

In this paper, we focus on  the transmission properties of quantum networks,
generalizing the original works of the 80's \cite{Buttiker1,Doucot}.  This
work is motivated by
recent conductance measurements of normal metallic networks   etched on a
2D GaAs/GaAlAs
electron gas \cite{Naud_T3}.  Remarkably, for the particular ${\cal T}_3$
network
shown in fig.~\ref{figure1},  the  magnetoresistance presents large
$\phi_0$-periodic oscillations  which are barely visible for a more
conventional
geometry like the square lattice.  This is the first time that $\phi_0$
oscillations are observed in a macroscopic system where, in principle, ensemble
average due to a finite coherence length is expected to destroy them.

The experimental study of the  ${\cal T}_3$ network has been motivated by the
recent prediction of a new type of magnetic  field induced localization.
Indeed,
it has been shown, in a tight-binding approach, that when the flux $\phi$ per
elementary plaquette equals $\phi_0/2$ (half-flux), the electron motion is
completely confined inside the so-called AB cages \cite{Vidal} resulting
from a subtle quantum
interference effect.
This surprising phenomenon has first been experimentally observed
in  superconducting  (${\cal T}_3$) networks\cite{Pannetier}, where it was
found that
the critical current almost vanishes at $\phi=\phi_0/2$. The standard mapping
between the Ginzburg-Landau theory and the tight-binding problem
\cite{Alexander}
actually allows one to relate this current to the energy band curvature,
predicting a zero
critical current at half-flux.
However, it is interesting to know whether this localization effect still
exists
in normal metallic networks and if it could be at the origin of the
oscillations discussed above \cite{Naud_T3}.

The aim of this paper is threefold. Firstly, we
provide a simple formalism allowing to calculate the transmission
coefficient of any
network made up of one-dimensional wires. Secondly, we concentrate on two
regular
structures, the square and the ${\cal T}_3$ networks and study the flux
dependence of
the transmission coefficient which is reminiscent of the butterfly-like
structure
of the tight-binding spectrum. We then consider the influence of elastic
disorder that we
model by a distribution of the wire lengths.
We show that the ${\cal T}_3$ network exhibits $\phi_0$-periodic
oscillations  which are
robust with respect to disorder and which are much larger than those
observed in the
square network. We also discuss the crossover from a ballistic (in the pure
case)
to a disorder dominated behaviour, revealed by the emergence of
$\phi_0/2$-periodic oscillations reminiscent of the weak localization regime.
This model gives a strong support to the interpretation of the above-mentionned
experiment \cite{Naud_T3} in terms of the AB cages.

\bigskip

We consider a graph made up of $N$ nodes and connected to $N_{in}$ wires
(also called channels)
defining the input reservoir and to $N_{out}$  wires defining
the outgoing reservoir (see fig.~\ref{figure1}). In the Landauer approach, the
two-terminal conductance is proportional to the total transmission
coefficient defined by~:
%
%
\be
T=\sum_{i,j} |t_{ij}|^2
\mbox{,}
\label{transdef}
\ee
%
%
where $i \in [1,N_{in}]$ denotes the $i^{th}$ input channel and
$j \in [N-N_{out}+1,N]$ denotes the $j^{th}$ output channel. This
coefficient is the sum
of each individual transmission coefficient obtained by injecting a
wavepacket in
the $i^{th}$ channel. We emphasize that actually eq. (\ref{transdef})
assumes that there is no phase relationship between  the different input
channels \cite{Imrybook}.
%
%
%
%
Let us  consider an incoming wavefunction in the $i^{th}$ channel defined by~:
%
%
\be
\psi(x) = e^{-i k x}+r_{ii} e^{i k x}
\mbox{,}
\ee
%
%
where $r_{ii}$ is the reflexion coefficient in this wire. We need to determine
the transmission coefficient $t_{ij}$ giving the probability for the wave
packet
to outgo into the $j^{th}$ channel. Therefore, we first solve the Schr\"odinger
equation  on each bond whose extremities are denoted by $\alpha$ and
$\beta$. The
corresponding eigenfunctions are simply given by~:
%
%
\be
\psi_{\alpha\beta}(x)={e^{-i \gamma_{\alpha x}} \over \sin k l_{\alpha\beta}}
\left[ \psi_\alpha \sin k(l_{\alpha\beta}-x) + e^{i \gamma_{\alpha \beta}}
\psi_\beta \sin k x
\right]
\mbox{,}
\ee
%
%
where $\gamma_{\alpha \beta} = (2 \pi / \phi_0)\int_\alpha ^\beta {\bf
A}.\mbox{d}{\bf l}$
is the circulation of the vector potential between $\alpha$ and $\beta$,
$k$ is the wave
vector related to the eigenenergy by~: $E(k)=\hbar^2 k^2 / 2m$, $x$ is the
distance
measured from the node
$\alpha$ and $l_{\alpha\beta}$ is the length of the bond $(\alpha,\beta)$.
The current
conservation on each internal node of the network (not connected to
reservoirs) is
satisfied if~:
%
%
\be M_{\alpha\alpha} \psi_\alpha +
\sum_{\langle \alpha,\beta \rangle} M_{\alpha\beta} \psi_\beta= 0
\mbox{,}
\label{inside}
\ee
%
%
where $M$ is a ($N \times N$) matrix whose elements are given by:
%
%
\be M_{\alpha \alpha} = \sum_{\langle \alpha,\beta \rangle} \cot k l_{\alpha
\beta}\ \ , \ \ M_{\alpha \beta} = - \frac{e^{i \gamma_{\alpha \beta}}} {\sin k
l_{\alpha \beta}}
\mbox{.}
\label{matrixM}
\ee
%
%
The symbol $\langle \alpha,\beta \rangle$ indicates that the sums extend to all
the nodes $\beta$ connected to the node $\alpha$. In addition, the
off-diagonal element
$M_{\alpha \beta}$ is non zero only if the nodes $\alpha$ and $\beta$ are
connected by a bond. Consider now the case where the current is injected in the
channel $i \in [1,N_{in}]$. The current conservation at this node writes~:
%
%
\be
 M_{ii} \psi_i +\sum_{\langle i ,\beta \rangle} M_{i\beta}\psi_\beta=
i(1-r_{ii})
\mbox{.}
\label{in}
\ee
%
%
For each node $j \in [N-N_{out}+1,N]$, one also has~:
%
%
\be
 M_{jj} \psi_j +\sum_{\langle j ,\beta \rangle} M_{ j \beta }\psi_\beta= -i
t_{ij}
\mbox{.}
\label{out}
\ee
%
%
Finally, for $i \in [1,N_{in}]$ and $j \in [N-N_{out}+1,N]$, the continuity
of the
wavefunction reads~:
$\psi_i=1+r_{ii}$ and $\psi_j=t_{ij}$. The equations
(\ref{inside},\ref{in},\ref{out}) constitute a $(N \times N)$ linear system
\cite{Smilansky}
from which $t_{ij}=\psi_j$ ($j \in [N-N_{out}+1,N]$) can be calculated. The
total
transmission coefficient  is finally obtained from eq.(\ref{transdef}) by
considering the $N_{in}$ input channels.\\

We now apply this formalism to  the case of regular networks where all the
bonds
have identical length $l$ so that the transmission coefficient $T(k,f)$ is a
periodic function of the wave vector $k$ with  period $2 \pi /l$ and  a
periodic
function of the reduced flux $f=\phi/\phi_0$ with  period $1$.
In principle, the $k$-dependence of the  transmission coefficient can be probed
experimentally, if the wave vector $k$ is  well defined, {\it i.e.} if the
energy
of injected electrons is well controlled. Several factors like finite
temperature
or  finite bias contribute to broaden this energy. This  can be taken into
account
by giving a finite width $\Delta k$ to the Fermi wave vector of the
incoming  wave packet.
 For example, in ref.  \cite{Nordita}    the conductance of a single ring was
measured and it was found that the phase of the AB oscillations could be
varied by
tuning the gate voltage, and thus the Fermi energy. One may
therefore conclude that the width $\Delta k$ is  smaller than the period $2 \pi
/l$. These oscillations are very well  described by a Landauer
single-channel formalism, assuming that the ring is assymetric, {\it i.e.}
the two
arms have a different length \cite{Buttiker1,Nordita}.

\begin{figure}
\centerline{ \epsfxsize=8cm
\epsffile{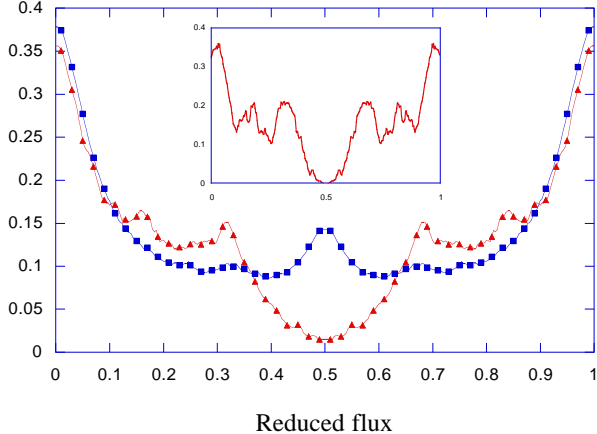}}
\vspace{2mm}
\caption{Averaged transmission coefficient $\langle T(k,f) \rangle_k /
N_{in}$  as a
function of the reduced flux for a $(8\times 8)$ square lattice (square)
and a piece
of the ${\cal T}_3$ network (triangle)  with 75 sites. Input and output
channels are
connected as displayed in fig.~\protect{\ref{figure1}}. Inset~: Averaged
transmission
coefficient for the
${\cal T}_3$ network with one input channel at the center of the network.}
\label{figure2}
\end{figure}

For a given $k$, the flux dependence of $T(k,f)$
has a rich  structure which is reminiscent  of the complexity of the associated
tight-binding spectrum.  Here, for simplicity, we have chosen to average the
transmission coefficient over a period
$k \in [0,2 \pi /l]$. The flux dependence of the average transmission $\langle
T(k,f) \rangle_k$ is shown  in fig.~\ref{figure2} for the square and
${\cal T}_3$ networks.
One clearly observes a few  peaks in the transmission for particular values
of the reduced
flux~: $f=1/2,1/3$ for the square lattice and $f=1/3,1/6$ for the
${\cal T}_3$ lattice.
One can simply understand this structure by invoking the extended nature of the
corresponding eigenstates that are Bloch waves with a spatial period
proportional to the
denominator of $f$ \cite{resolution}. Due to the existence of the AB cages, the
transmission coefficient is minimum at $f=1/2$ for the ${\cal T}_3$ network
but,
surprisingly, it is not exactly zero. This is due to the existence of
dispersive
edges states
\cite{Vidal_T3_big} that are able to carry current even for $f=1/2$.
Therefore,
$T$ converges toward a finite value for the ${\cal T}_3$ network when the
system size (and
$N_{in}$) increases, whereas $T\sim N_{in}$ in the square lattice.
However, when one injects current in the bulk of the sample, the
transmission completely
vanishes for this flux. (see inset fig.~\ref{figure2}).
This study shows that the cage effect, originally predicted in a
tight-binding model, also
arises in a ${\cal T}_3$ network made up of one-dimensional ballistic  wires.
%
%
%
%
%
%

We now consider the case of  disordered networks, the motivation being to see
whether the cage phenomenon persists in such a situation.  Disorder can be
introduced in several ways (randomly distributed pointlike scatterers, or more
generally, random elastic scattering matrix along the bonds).
Here, in order to simulate random phase shifts on each bond, we consider a
geometrical
disorder defined  by a random modulation of the  wire lengths while keeping
the same
connectivity. Denoting by
$\Delta l$  the amplitude of the length fluctuations, the relevant
dimensionless
parameter to characterize the strength of the disorder is the quantity $k
\Delta l$ and thus
explicitly depends on the energy.   Note that the incommensurability
between the different
lengths breaks the periodicity of $T$ with respect to $k$.
This type of disorder also provides a distribution of areas of width $2
l\Delta l$ so that
the oscillations are expected to  disappear after about $l /\Delta l$ periods.
In the following, we will focus on situations where  $\Delta l/l \ll 1$ and
$k \:l \gg 1$ so that the case $k \Delta l \sim 1$ may be reached without a
sizeable
dispersion of the areas. Thus, we will not modify the bond lengths in the
phase factor
$e^{i\gamma_{\alpha\beta}}$ and the periodicity with respect to the reduced
flux will be
conserved.

\begin{figure}
\centerline{\epsfxsize=8cm
\epsffile{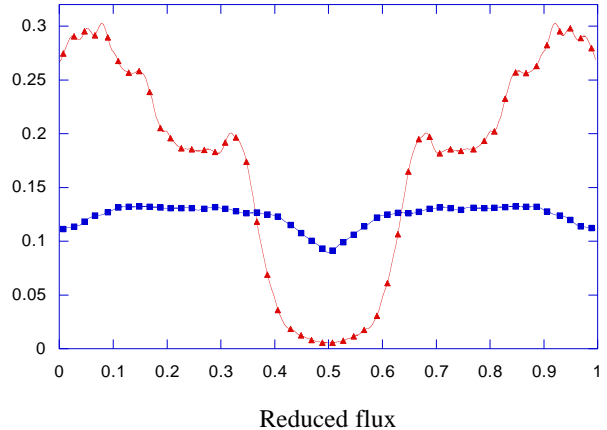}}
\vspace{2mm}
\caption{Transmission coefficient $\langle T(k,f) \rangle_{dis.}/N_{in}$
averaged over 50
configurations of disorder for $k l=\pi/3$  and $k\Delta l = 1.47$ as a
function of the reduced
flux.}
\label{figure3}
\end{figure}

For a given realisation of disorder, $T(k,f)$ exhibits a $\phi_0$-periodic
complex
structure which is a signature of the interference pattern through the
network. In particular,
the transmission extremely sensitive to $k$.
However, experimentally, there is always a finite phase coherence length
$L_\varphi$.
Therefore, a two-dimensional network of typical linear size $L$ must be
considered as a set of
$(L/L_\varphi)^2$ regions without phase relationship. This provides a
natural averaging
mechanism over disorder realisations. Thus,  we have chosen to study the
disorder averaged
transmission coefficient $\langle T(k,f) \rangle_{dis.}$ whose variations
versus the reduced flux are displayed in fig.~\ref{figure3} for fixed
$k$ and disorder strength.
It is clearly seen that for the square network, the periodicity of
$\langle T(k,f) \rangle_{dis.}$ with respect to the magnetic flux is no
longer $\phi_0$ but
$\phi_0/2$.  The $\phi_0$-periodic oscillations have been washed out since
they do not have a
given  phase. By contrast, the $\phi_0/2$-periodic oscillations are still
present since they
are related to phase coherent pairs of time-reversed trajectories according
to the
weak-localization picture.  For the ${\cal T}_3$ network, the transmission
coefficient remains
$\phi_0$-periodic with a large amplitude. This strongly suggests that the
cage effect
(which locks the phase of the oscillations) survives for this strength of
disorder.

%
%
%
%
%
%

\begin{figure}
\vspace{2mm}
\centerline{ \epsfxsize=7cm
\epsffile{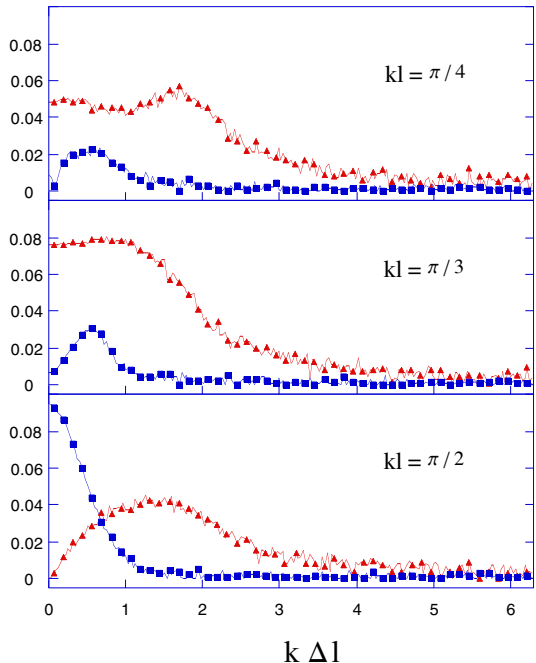}}
\vspace{2mm}
\caption{Variation of $\langle |\tilde T(k,1)| \rangle_{dis.}/N_{in}$
versus disorder for
different values of $k$ after averaging over 50 configurations.}
\label{figure4}
\end{figure}

For a finer analysis, it is interesting to compute the discrete Fourier
transform of $T$ defined by~:
%
%
\be
\tilde T(k,\omega)= {1\over n} \sum_{j=0}^{n-1} T(k,j/n) e^{i 2 \pi\omega j/n}
\mbox{,}  \:\:\: \omega \in [0,n-1]
\mbox{,}
\ee
%
%
where $n$ is the number of sampled values of $f$.
Figure \ref{figure4} displays $\langle |\tilde T(k,1)| \rangle_{dis.}$ as a
function of the
disorder strength
$k \Delta l$ for different values of $k$.
It shows that, when disorder is increased, $\langle |\tilde T(k,1)|
\rangle_{dis.}$
persists much longer for the ${\cal T}_3$ network than for the square
network. We are
thus led to conclude that the cage effect is robust with respect to disorder.
Note that for weak disorder, $\langle |\tilde T(k,1)| \rangle_{dis.}$
depends on $k$ but
this dependence vanishes for $k \Delta l \gtrsim 2$.
We strongly believe that this result  explains why a $\phi_0$-periodic
conductance is observed
experimentally  for the ${\cal T}_3$ network while it is not for the square
lattice \cite{Naud_T3}.

\begin{figure}
\centerline{\epsfxsize=7cm
\epsffile{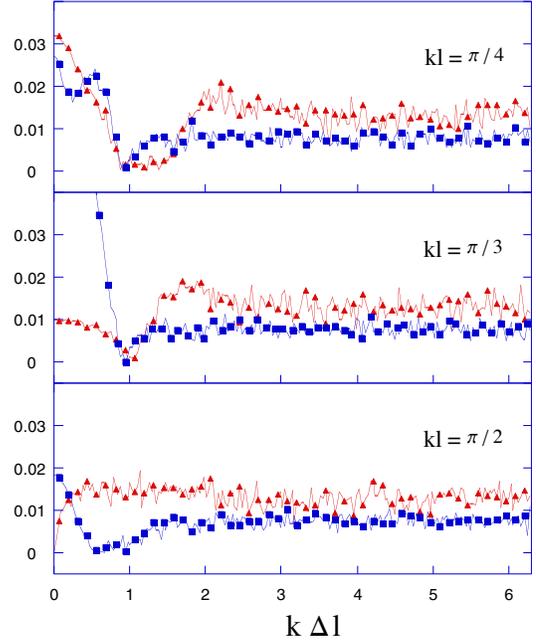}}
\vspace{2mm}
\caption{Amplitude of the second harmonic $\langle |\tilde T(k,2)|
\rangle_{dis.}/N_{in}$
versus disorder for different values of $k$ after averaging over 50
configurations.}
\label{figure5}
\end{figure}

The behaviour of $\langle |\tilde T(k,2)| \rangle_{dis.}$ is shown in
fig.~\ref{figure5}. It
is interesting to see that this harmonic becomes quickly dominant for both
networks and
remains constant for  $k \Delta l \gtrsim 2$. The value of this constant
depends on the system
size and converges to zero for the infinite lattice. Nevertheless, we leave
the precise
analysis of this scaling for further studies.

Finally, it should  be stressed that, experimentally,
$\langle |\tilde T(k,2)|
\rangle_{dis.}$ is further reduced  by a factor $e^{-2 L / L_\varphi}$ due
to a finite
coherence length $L_\varphi$, while the $\phi_0$ contribution is only
reduced by a factor
$e^{-L /L_\varphi}$, $L$ being the perimeter of an elementary plaquette
\cite{Aronov87}.

%
%
%
%
%
%

In conclusion, we have used a simple and general formalism to calculate the
transmission coefficient of any network made up of single-channel quantum
wires.
This coefficient can be simply expressed in terms of a connectivity-like
matrix.
We have used this formalism to study the AB cage phenomenon in the ${\cal T}_3$
network and we have shown that this effect is  robust to a moderate amount of
elastic disorder. As a consequence, the AB oscillations with period
$\phi_0$ persist in the
infinite ${\cal T}_3$ networks whereas they vanish in the square lattice.

We acknowledge  H. Bouchiat, P. Butaud, R. Deblock, G. Faini, D. Mailly, R.
Mosseri, C.
Naud and B. Reulet for useful discussions.

%
%
%
%
%
%


\begin{thebibliography}{99}

\bibitem{Webb}S. Washburn and R. Webb, Adv. Phys. {\bf 35}, 375 (1986).

\bibitem{Buttiker1} M. B\"uttiker, Y. Imry and M. Ya. Azbel, Phys. Rev. A
{\bf 30}, 1982
(1984).

\bibitem{Buttiker2} M. B\"uttiker,  Y. Imry, R. Landauer and S. Pinhas,
Phys. Rev. B {\bf 31},
6207 (1985).

\bibitem{Aronov87} A.G. Aronov and Yu.V. Sharvin, Rev. Mod. Phys. {\bf 59},
755 (1987).

\bibitem{Pascaud99} M. Pascaud and G. Montambaux, Phys. Rev. Lett. {\bf
82}, 4512 (1999).

\bibitem{Doucot} B. Dou{\c c}ot and R. Rammal, J. Physique (Paris) {\bf
48}, 941 (1987).

\bibitem{Naud_T3} C. Naud, G. Faini, D. Mailly and B. Etienne preprint~:
cond-mat/006400.

\bibitem{Vidal} J. Vidal, R. Mosseri and B. Dou\c cot, Phys. Rev. Lett.
{\bf 81}, 5888 (1998).

\bibitem{Pannetier}C.C. Abilio {\it et al.}, Phys. Rev. Lett. {\bf 83},
5102 (1999).

\bibitem{Alexander}  P.G. de Gennes, C.R. Acad. Sci. Ser. B {\bf 292}, 9
and 279 (1981),
S. Alexander, Phys. Rev. B {\bf 27}, 1541 (1983).

\bibitem{Imrybook} Y. Imry, {\it Introduction to mesoscopic physics},
Oxford University Press
(1997).

\bibitem{Smilansky} T. Kottos and U. Smilansky, Ann. Phys. {\bf 274}, 76
(1999).
\bibitem{resolution} Note that since the peak resolution is set by the
size of the sample, one would observe higher order peaks for much bigger
systems.

\bibitem{Nordita} S. Pedersen {\it et al.}, Phys. Rev. B {\bf 61}, 5457 (2000).

\bibitem{Vidal_T3_big}J. Vidal {\it et al.}, in preparation.

\end{thebibliography}
\end{document}